\documentclass[fleqn,usenatbib]{mnras}

\usepackage[T1]{fontenc}
\usepackage{ae,aecompl}

\usepackage{graphicx}	
\usepackage{amsmath}	
\usepackage{amssymb}	

\title[Is PSR J0855$-$4644 responsible for 1.4 TeV bump?]{Is PSR J0855$-$4644 responsible for the 1.4 TeV electron spectral bump hinted by DAMPE?}

\author[Y. Bao et al.]{
Yiwei Bao,$^{1}$
Yang Chen,$^{1,2}$\thanks{E-mail: ygchen@nju.edu.cn}
Siming Liu$^{3,4}$\thanks{E-mail: liusm@pmo.ac.cn}
\\
$^{1}$Department of Astronomy, Nanjing University, 163 Xianlin Avenue, Nanjing 210023, China\\
$^{2}$Key Laboratory of Modern Astronomy and Astrophysics, Nanjing University, Ministry of Education, Nanjing, China\\
$^{3}$Key Laboratory of Dark Matter and Space Astronomy, Purple Mountain Observatory, Chinese Academy of Sciences, Nanjing 210034, China\\
$^{4}$School of Astronomy and Space Science, University of Science and Technology of China, Hefei 230026, China\\
}

\date{Accepted XXX. Received YYY; in original form ZZZ}

\pubyear{2020}

\begin{document}
\label{firstpage}
\pagerange{\pageref{firstpage}--\pageref{lastpage}}
\maketitle

\begin{abstract}
DAMPE observation on the cosmic ray electron spectrum hints a narrow excess at $\sim 1.4$ TeV. Although the excess can be ascribed to dark matter particles, pulsars and pulsar wind nebulae are believed to be a more natural astrophysical origin: electrons injected from nearby pulsars at their early ages can form a bump-like feature in the spectrum due to radiative energy losses. In this paper, with a survey of nearby pulsars, we find 4 pulsars that may have notable contributions to $\sim1.4$ TeV cosmic ray electrons. Among them, PSR J0855$-$4644 has a spin down luminosity more than 50 times higher than others and presumably dominates the electron fluxes from them. X-ray observations on the inner compact part (which may represent a tunnel for the transport of electrons from the pulsar) of PWN G267.0$-$01.0 are then used to constrain the spectral index of high energy electrons injected by the pulsar. We show that high-energy electrons released by PSR J0855$-$4644 could indeed reproduce the 1.4 TeV spectral feature hinted by the DAMPE with reasonable parameters.
\end{abstract}




\section{Introduction}

Due to severe energy losses, high energy electrons have a short lifetime in the ISM, and cosmic ray electrons are thus believed to be vital probe of nearby dark matter \citep{2005PhR...405..279B} and astrophysical sources \citep{1970ApJ...162L.181S,1995A&A...294L..41A}. The e$^\pm$ excesses \citep[see e.g.,][]{2014PhRvL.113l1102A,2019PhRvL.122d1102A} have inspired a number of works discussing their possible origins. The excesses can be attributed to dark matter \citep[see e.g.,][]{2008PhRvD..78j3520B,2020MNRAS.491..965G} and/or astrophysical sources \citep[see e.g.,][]{2009PhRvL.103e1104B,2009PhRvL.103k1302S,2009JCAP...01..025H}. The DAMPE experiment has reported its high energy resolution measurement of the cosmic ray electron flux up to 4.6 TeV \citep{2017Natur.552...63D}, hinting at a narrow excess at $\sim 1.4$ TeV. The local significance of the narrow excess, according to \citet{2018PhRvD..97i1701H}, is $\sim3.7 \sigma$. Although the bump is marginally significant and CALET found no significant line-like feature in the same energy range \citep[see e.g.,][]{2018PhLB..780..181F,2018PhRvD..98l3008J,2018PhRvL.120z1102A}, it attracts wide attention and a certain number of models have been proposed to discuss the possible origin of it \citep[see e.g.,][]{2018JCAP...05..053N,2018JHEP...02..121A,2018JCAP...05..053N,2018PhRvD..97i1701H,2018PhRvD..98c5025L,2019JCAP...10..063C,2019PDU....2600333B,2019MNRAS.486L..85C,2020MNRAS.491..965G}. 

Mature pulsars (with age $\ge 10^5$ yr) which are sufficiently close to Earth are believed to be the plausible contributor of bump-like features in the cosmic ray electron spectrum. The nonthermal electron spectra of young pulsars are usually hard \citep{2008ApJ...682.1166L}. \citet{2009PhRvD..80f3005M} argued that the existence of bumps at high energies strongly suggest that the feature is produced by pulsars. Pulsars are ideal factories for high-energy electrons and positrons. On the surface of pulsar, the seed electrons are pulled out of the crust by the intense electric field (induced by the spin of the pulsar), and accelerated. Moving along magnetic field, the electrons emit photons via curvature radiation, and numerous e$^{\pm}$ pairs are produced as the emitted photons interact with the magnetic field \citep{1994SSRv...68..363V}. The electrons are then transported outwards in forms of ultra-relativistic pulsar wind, and accelerated at the termination shock or even earlier. Since the lifetime of pulsars is much longer than those of supernova remnants, pulsars of ages $\sim 10^2$ kyr are expected to interact with interstellar medium (ISM) directly. Double bow shocks form when the pulsar wind interacts with the incident ISM \citep[e.g.,][]{1996ApJ...469..715C,2003Sci...301.1345C,2006ARA&A..44...17G}. The accelerated electrons are expected to escape from the tail of the bow shock \citep{2018MNRAS.480.5419B}, traveling a long distance so as to arrive at the Solar system after diffusion in the ISM. 


In this paper, we demonstrate that the PSR J0855$-$4644 (as well as its PWN G266.97$-$1.00) is a promising electron contributor for 1.4 TeV spectral bump. The criterions for contributor candidates are listed in \S 2, the spectral energy distribution (SED) of the PWN G266.97$-$41.00 is explained in \S 3, and  its contribution to cosmic ray electrons is calculated in \S 4.

\section{candidate selection}
First of all, we constrain the ages of candidates which may be responsible for the 1.4 TeV bump.  As is shown in \citet{2018APh...102....1L}, the Inverse Compton (IC) off optical photons is severely suppressed for electrons above 100 GeV, we therefore approximate the energy loss term to be \citep{1995PhRvD..52.3265A}
\begin{equation}
P(\gamma)=b_0 \gamma^2=5.2 \times 10^{-20} \frac{u_{\rm B}+u_{\rm CMB}+u_{\rm FIR}}{1 \rm\ eV\ cm^{-3}} \gamma^2, 
\end{equation}
where $u_{\rm CMB}$ and $u_{\rm FIR}$ represent the energy density of cosmic microwave background (CMB) radiation and far infrared (FIR) photons, respectively, and $u_{\rm B}=B^2/(8\pi)$ represents the energy density of magnetic field (with a strength $B$). The photon energy density in the Solar vicinity is $u_{\rm FIR}=0.4$ eV cm$^{-3}$, and $u_{\rm CMB}=0.25$ eV cm$^{-3}$ \citep{2002cra..book.....S}, and the mean magnetic field near the Sun is $B\approx 6\ \mu$G \citep[see e.g.,][]{2015A&ARv..24....4B}. The time required for an electron to cool down from an infinity energy to $1.4$ TeV is $\tau_{\rm CD} \approx 1.4 \times 10^5$ yr. Since half (assuming braking via magnet dipole radiation) of the spin-down energy is released within its initial spin-down timescale $\tau_0$ (which is usually much less than age of the pulsar), the electrons injected till $\tau_0$ will form a bump in the electron spectrum at time $\approx \tau_{\rm CD}$ if the injection spectrum is hard (with the power index significantly smaller than 2, \citeauthor{2018PhRvD..97i1701H}\,2018). Therefore the age of the contributor candidates of the $1.4$ TeV feature should be $T_{\rm age} \approx \tau_{\rm CD}\approx1.4 \times 10^5$ yr.


The actual age of a pulsar, however, is hard to determine if there is no record of its parent supernova in the literature. The characteristic age of the pulsar $\tau_{\rm c}$ can only give a crude estimate to the actual age $T_{\rm age}$ \citep[$\tau_{\rm c}$ usually overestimates $T_{\rm age}$, see e.g.,][]{2006ARA&A..44...17G}, and hence all pulsars with characteristic age 130--200 kyr are taken for the survey. Following Model DC in \citet{2017PhRvD..95h3007Y}, we adopt the diffusion coefficient in the ISM $D_{\rm ISM}=4.2 \times 10^{28} (E/4\ {\rm GeV})^{\delta}$ (with $\delta=0.588$), and the diffusion length scale of $\sim 1.4$ TeV electrons would be \citep{1995PhRvD..52.3265A}
\begin{equation}
R_{\rm diff}(\gamma,t)=2\sqrt{D_{\rm ISM}(\gamma)t\frac{1-(1-\gamma/\gamma_{\rm max})^{1-\delta}}{(1-\delta)\gamma/\gamma_{\rm max}}}\approx {2}\ {\rm kpc},
\end{equation}
where $\gamma_{\rm max}=(b_0T_{\rm age})^{-1}$. So candidates are expected to be located within {2} {\rm kpc} kpc from Earth.

In the pulsar catalog\footnote{https://www.atnf.
csiro.au/research/pulsar/psrcat} \citep{2005AJ....129.1993M}, we find 4 pulsars/PWNe with a characteristic age of 130--200 kyr and a distance within {2} kpc. The distances and the spin-down luminosities of these pulsars are listed in \autoref{tab:Ld}. There are some uncertainties about the distance of PSR J0855$-$4644 since no accurate parallax measurements are available. The dispersion measure gives a distance of $\sim 6$ kpc \citep{2005AJ....129.1993M}. However, based on the measurements of the intervening hydrogen column density $N_{\rm H}$, \citet{2013A&A...551A...7A} set an upper limit 900 pc to J0855$-$4644 considering the $N_{\rm H}$ value towards the pulsar is less than that towards the rim of the SNR Vela Junior. In this paper, we assume PSR J0855$-$4644 is at a distance of 900 pc following \citet{2017A&A...597A..75M}. Due to its remarkably high spin-down luminosity at present (two orders of magnitude higher than those of other pulsars), PSR J0855$-$4644 is the most promising contributor (quantitative comparison can be found in \S 4).

\begin{table*}
	\centering
	\caption{Basic parameters of the selected pulsars}
	\label{tab:Ld}
	\begin{tabular}{lccr} 
		\hline
		
pulsar & Distance & Spin-down luminosity at present & Characteristic age \\
&(kpc) & ($10^{34}$ erg s$^{-1}$)& ($10^5$ yr) \\
		\hline
J0157+6212	&	1.79	&	0.057	&	1.97	\\
J0855-4644 	&	0.9 (assumed)		&	110	&	1.41	\\
J0954-5430	&	0.43	&	1.6	&	1.71	\\
J1301-6310	&	1.46	&	0.76	&	1.86	\\
		\hline
	\end{tabular}
\end{table*}

\section{the synchrotron radiation of the PWN G266.97$-$1.00}

Here we constrain the electron powerlaw index $\alpha$, and the cutoff Lorentz factor $\gamma_{\rm cut}$ with the SED of PWN G266.97$-$1.00. We assume $\alpha$ and $\gamma_{\rm cut}$ are independent of time i.e., the electrons injected in early age (which could account for the 1.4 TeV bump) and the electrons injected recently (which forms PWN G266.97$-$1.00) have the same $\alpha$ and $\gamma_{\rm cut}$. 

\begin{center}
\begin{figure}
\includegraphics[scale=0.5]{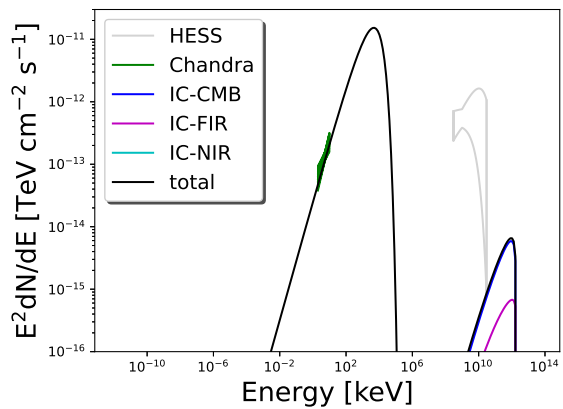}
\caption{SED of inner compact part of the PWN G267.0$-$01.0. The Chandra data are taken from \citet{2017A&A...597A..75M}, and the HESS data are taken from \citet{2018A&A...612A...1H}. The HESS data are used as an upper limit in view of possible pollution from the overlapping SNR Vela Junior.}
\label{fig:SED}
\end{figure}
\end{center}

Located inside the PWN, PSR J0855$-$4644 is one of the most powerful pulsars lying nearby \citep{2013A&A...551A...7A}. The contributed flux from PSR J0855$-$4644 to the observed TeV e$^{\pm}$s has been proposed for years \citep{2013A&A...551A...7A}. The PWN lies projectionally near the south-eastern rim of the SNR Vela Junior \citep{2017A&A...597A..75M}. As seen in radio, G266.97$-$1.00 is a bow shock PWN \citep{2018MNRAS.477L..66M} with a faint trailing tail. Since the faint tail has blent into the background noise, the tail might be much longer than seen. 
Chandra X-ray observation reveals a compact part with double torus/jet morphology ($\sim 10''$, or $\sim 0.04$ pc at a distance of 900 pc) in the very vicinity of the pulsar \citep{2017A&A...597A..75M}, while XMM-Newton X-ray observation find a much larger diffuse outer nebula of $\sim 1'$ \citep{2013A&A...551A...7A}. The origin of jet-like features in bow shock PWNe have been detailedly studied in recent years; \citet{2019MNRAS.490.3608O} show that the jets can be explained as the major tunnel of the escaping high energy electrons. The projective distance from the pulsar to the bow shock apex is $\sim 20''$ ($\sim 0.3$ pc) in the radio observation \citep{2018MNRAS.477L..66M}. The standoff distance, however, is smaller than the projective value by about 10\%, 20\%, and 50\% for inclination angles $60^\circ$, $45^\circ$, and $20^\circ$, respectively \citep{2017ApJ...842..100N}, and therefore, the double torus/jet sturcture could be thought to lie within the unshocked pulsar wind\footnote{\citet{2017A&A...607A.134C} show that the magnetic field can dissipate far before reaching the termination shock. The Chandra observation on PWN G266.97$-$1.00 is consistent with the model proposed by \citet{2017A&A...607A.134C}, since the inner PWN is much smaller than the size of the termination shock.}. Here we approximate that the velocity of the jet is $\sim c$. The compact nebula is in dynamic equilibrium: the pulsar keeps injecting electrons, and the electrons are transported outwards through the double torus/jet structure. The advection timescale of the system, is $T_{\rm conv}=l_{\rm jet}/c\approx0.07$ yr, where $l_{\rm jet}$ is the length of the jet-like structure ($\sim0.02$ pc, half of the size of X-ray morphology). While the electron transport is determined by advection in the compact nebula (within $\sim 10''$), diffusion also participates in the tranport in the outer diffuse nebula ($\sim 1'$). To avoid introducing new parameters describing the diffusion, we only explain the SED of the electrons in the inner compact nebula (or the escaping tunnel). The electron spectrum is calculated by solving the equation
\begin{equation}
\frac{\partial N}{\partial t}=\frac{\partial }{\partial \gamma}\left[P(\gamma)N(\gamma,t)\right]+Q_{\rm i}(t)\gamma^{-\alpha}H({\gamma_{\rm cut}-\gamma}),
\end{equation}
where $N(\gamma,t)$ represents the differential electron number, $\gamma$ the Lorentz factor of the electrons, $H$ the Heaviside step function, and $Q_{\rm i}(t)$ the normalization constant which is determined by the injection rate of the pulsar. We assume that almost all of the spin-down energy is converted into electron energy i.e.,
\begin{equation}
\int_1^{\gamma_{\rm cut}}Q_{\rm i}(t)\gamma^{-\alpha}\gamma\,d\gamma=L(t)=\frac{L_0}{(1+t/\tau_0)^{(n+1)/(n-1)}}, 
\end{equation} 
where $L(t)$ is the spin-down luminosity at time $t$, {$\alpha$ the injection power law index,} $L_0$ the initial spin-down luminosity, $n$ the braking index, and $\tau_0=2\tau_{\rm c}/(n-1)-T_{\rm age}$. To fit the X-ray flux, the magnetic field in the compact part is adopted to be $B_{\rm compact}=30\ \mu$G. The code is run from $T_{\rm age}-0.07$ yr to $T_{\rm age}$, the resulting SED of the photons is plotted in \autoref{fig:SED}, and the parameters are listed in \autoref{tab:par}, where $T_{\rm OPT}$, $T_{\rm FIR}$, $T_{\rm CMB}$ are the black body temperature of optical, FIR, and CMB photons, respectively. The fitted power law index of the electrons that are accelerated by the PWN G267.0$-$01.0 is 1.1, which is hard enough to account for the narrow bumped hinted by DAMPE. Due to its limited angular resolution, the HESS observation has not resolved the PWN, and the $\gamma$-ray data may be polluted by the overlapping SNR Vela Junior, and therefore we only use the HESS data as an upper limit in the above fitting. Given the hard spectrum, the cutoff energy $\gamma_{\rm cut}$ is well constrained by the convection time. The cutoff energy will decrease for a longer convection time.

\begin{table}
	\centering
	\caption{Fitting Parameters}
	\label{tab:par}
	\begin{tabular}{lcr} 
		\hline
		Quantity & Value & Comment\\
\hline
\multicolumn{3}{c}{Time independent Quantity}\\
\hline
$\gamma_{\rm cut}$	& $4 \times 10^9$				& Parameter \\
$\alpha$		& $1.1$					& Parameter \\
$T_{\rm CMB}$		& $2.73$ K					& Observed \\
$u_{\rm CMB}$		& 0.25 eV cm$^{-3}$				& Observed \\
$T_{\rm FIR}$		& $20$ K					& Parameter \\
$u_{\rm FIR}$		& 0.4 eV cm$^{-3}$				& Parameter \\
$T_{\rm OPT}$		& $5000$ K					& Parameter \\
$u_{\rm OPT}$		& 0.3 eV cm$^{-3}$				& Parameter \\
\hline
\multicolumn{3}{c}{Parameters for the inner compact part of PWN at present}\\
\hline
$T_{\rm conv}$		& 0.07 yr					& Parameter\\
$L_{\rm now}$			& $1.1 \times 10^{36}$ erg s$^{-1}$ 	& Observed \\
$B_{\rm compact}$		& $30\ \mu$G					& Parameter\\
\hline
\multicolumn{3}{c}{Parameters for the continuous injection in whole lifetime}\\
\hline
$n$			& 3						& Assumed  \\
$T_{\rm age}$		& 132 kyr 					& Parameter \\
$B$			& $6.66\ \mu$G					& Parameter \\
\end{tabular}
\end{table}


\section{Contribution to all-electron spectrum}
\begin{center}
\begin{figure}
\includegraphics[scale=0.5]{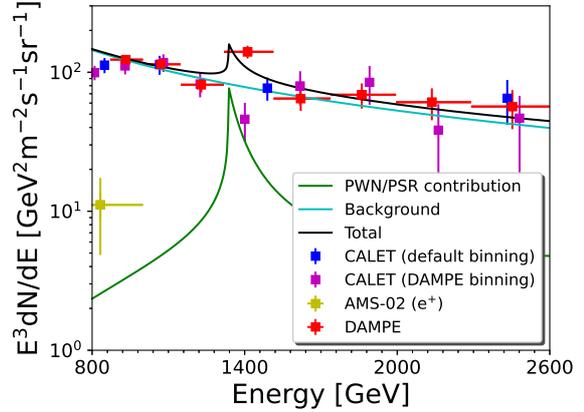}
\caption{The electron spectrum observed near Earth, fitted with the contribution from PSR J0855-4644. The background e$^\pm$s are contributed by other nearby sources such as SNRs. The DAMPE data are taken from \citet{2017Natur.552...63D}, CALET data from \citet{2018PhRvL.120z1102A}, and AMS positron data from \citet{2019PhRvL.122d1102A}.}
\label{fig:elec}
\end{figure}
\end{center}

We now calculate the electron distribution function {$f$} {by solving the equation}
\begin{equation}
\label{eq:dif}
\frac{\partial}{\partial t}f(\gamma,R,t)=\\ \frac{D_{\rm ISM}(\gamma)}{R^{2}}\frac{\partial}{\partial R}R^{2}\frac{\partial}{\partial R}f(\gamma,R,t)+\frac{\partial}{\partial \gamma}\left(Pf\right)+Q_{\rm i}(t)\gamma^{-\alpha}.
\end{equation}
{According to \citet{1995PhRvD..52.3265A}, the Green function for the \autoref{eq:dif} is ($\Delta t=T_{\rm age}-t$)}  
\begin{equation}
G(\gamma,R,\Delta t)=\frac{1}{\pi^{3/2}R^3}(1-b_0\Delta t \gamma)^{\alpha-2}\left(\frac{R}{R_{\rm diff}}\right)^3e^{-(R/R_{\rm diff})^2}.
\end{equation}
{Finally, $f$ can be obtained via}
\begin{equation}
f=\int_0^{T_{\rm age}} G(\gamma,R,\Delta t)Q_{\rm i}(t)\gamma^{-\alpha}\,{\rm d}\Delta t.
\end{equation}
The electron flux in the Solar vicinity contributed by PSR J0855$-$4644 is shown in \autoref{fig:elec}, with the parameters listed in \autoref{tab:par}. As seen in \autoref{fig:elec}, a bump in the electron SED appears at 1.4 TeV and matches the observation data\footnote{Since $\tau_0\sim 1$ kyr for many pulsars (such as the Crab pulsar), and $\lesssim 10$ kyr for most pulsars, much smaller than the age of mature pulsars, the injection can be approximated to be impulsive \citep[see e.g.,][]{2009PhRvD..80f3005M}. Hence, although we use a continuous injection rate, the resulting electron spectrum is similar to that stems from an impulsive injection.}.
This indicates that the electrons released from PSR J0855$-$4644 could account for the presence of the 1.4 TeV electron bump detected by DAMPE. 

\begin{center}
\begin{figure}
\includegraphics[scale=0.5]{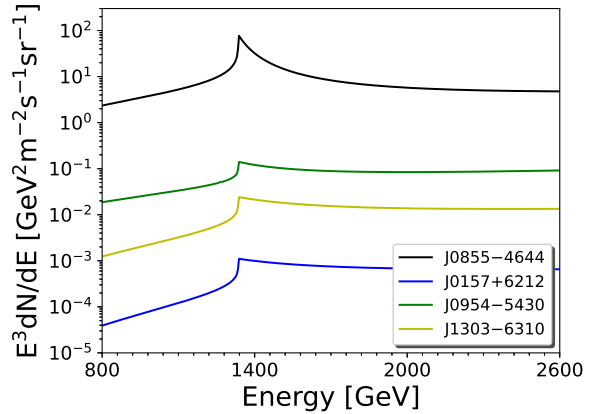}
\caption{The electron flux contributed by all listed pulsars.}
\label{fig:allpsr}
\end{figure}
\end{center}
\begin{center}
\begin{figure}
\includegraphics[scale=0.5]{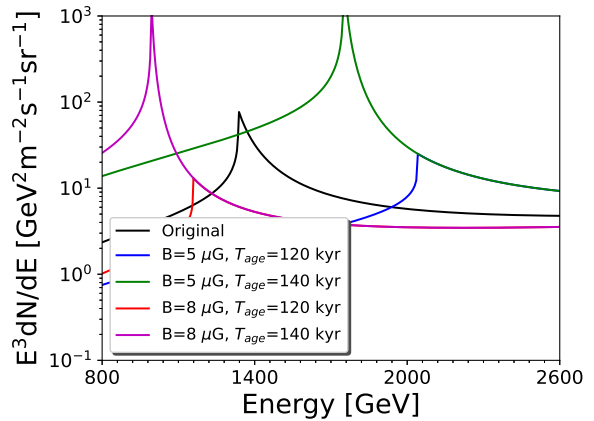}
\caption{The electron flux contributed by PSR J0855$-$4644 with various parameters.}
\label{fig:parvar}
\end{figure}
\end{center}

As plotted in \autoref{fig:allpsr}, we calculate the electron flux contributed by all the listed pulsars assuming $\alpha=1.1$, $\gamma_{\rm cut}=4\times10^9$ and $n=3$. The contribution from PSR J0855$-$4644 is much higher than those from other listed pulsars, indicating it as the most plausible contributor of the 1.4TeV electrons among the selected ones. In \autoref{fig:parvar}, we present the electron flux contributed by PSR J0855$-$4644 with various $B$ and $T_{\rm age}$. Assuming $n=3$, the total spin-down energy is calculated via $\tau_0$ i.e., $E_{\rm tot}=L_0\tau_0=L_{\rm now}(1+T_{\rm age}/\tau_0)^2\tau_0$, hence the electron flux increases quickly with the increase of $T_{\rm age}$.




\section{discussion}
The positron fraction in the electron flux contributed by pulsars via pair production is expected to be $\approx50\%$, so the positron flux near Earth may include the contribution from PSR J0855$-$4644 and other nearby pulsars and/or dark matter. Therefore the positron spectrum measured by the AMS-02 \citep{2019PhRvL.122d1102A} should be the upper limit of the positrons contributed by the pulsar. The AMS-02 reveals a soft positron spectrum at hundreds of GeV, while the pulsar is expected to contribute a hard positron spectrum, and therefore we only plot the last data point of the AMS-02 experiment in \autoref{fig:elec}. As can be seen, the predicted positron flux contributed by PSR J0855$-$4644 is much less than that is measured by AMS-02, and therefore the pulsar scenario does not contradict the measurement.

Due to uncertainty in the actual age of the pulsar, PSR J0855$-$4644 could potentially also account for other bump-like features in a few TeV. The anisotropy of signals can be used to distinguish different scenarios that could explain the electron spectrum observed. As was discussed, the smoking-gun signature of the pulsar/PWN scenario is anisotropy \citep{2013ApJ...772...18L}: the anisotropy of the electron flux contributed by a point source (pulsar) is much larger than that contributed by the diffuse dark matter. As is predicted in \citet{2018PhRvD..97i1701H}, the anisotropy is within the Fermi upper limit if the bump is contributed by a burst-like pulsar lying within 3 kpc, and further data can be measured in the near future with the help of LHAASO \citep[see e.g.,][]{2019IJMPD..2830022G}.

HAWC Observations of the Geminga PWN has confirmed that high-energy electrons are able to escape from the bow shock through diffusion \citep{2017Sci...358..911A}. The diffusion coefficient estimated is much smaller than the mean diffusion coefficient in the ISM. Similar phenomenon also occurs near many other pulsars \citep{2020PhRvD.101j3035D}. Such a slow diffusion can enhance the electron flux observed at earth \citep{2018ApJ...863...30F} implying a shorter age and/or larger distance than the model parameters derived above. However, due to the overlapping SNR Vela Junior, the existence of diffusive halo around PSR J0855$-$4644 has not been verified, and therefore we do not consider slow diffusion in this paper. 

\section{summary}

The DAMPE observation on the cosmic ray electron spectrum hints at a narrow excess at energy $\sim 1.4$ TeV. While dark matter lying nearby are usually considered to be the contributors of the electrons measured at Earth, pulsars and PWNe are also proposed to be promising astrophysical contributors. The electrons injected from nearby pulsar(s) at their early ages can be expected to form a bump in the spectrum at 1.4 TeV after diffusion. In this paper, after a survey of the nearby pulsars that are possible candidates to contribute a notable amount of electrons at 1.4 TeV, we select 4 pulsars in light of their characteristic ages and distances. We constrain the electron power law index and the cutoff Lorentz factor with the SED of PWN G266.97$-$1.00 on the assumption that they do not vary with time. We fit the SED, and use the same electron power-law and cutoff Lorentz factor to calculate the electron flux contribution. We consequently show that the electrons diffused from the PWN indeed display a bump-like distribution near 1.4 TeV, which may well match the DAMPE data. Meantime, we estimate the electron flux contributions from other selected pulsars and show that contribution from PSR J0855$-$4644 dominates over those from all other pulsars.



\section*{acknowledgments}
We would like to acknowledge Ruo-Yu Liu, Xiang-Dong Li and Pasquale Blasi for helpful comments. This work is supported by the National Key R\&D Program of China under grants 2017YFA0402600 and 2018YFA0404203, NSFC under grants 11773014, 11633007, 11851305, U1738122, {U1931204}, and 11761131007, and the International Partnership Program of Chinese Academy of Sciences under grant 114332KYSB20170008.

\section*{Data availability}
No new data were generated or analysed in support of this research.





\begin{thebibliography}{99}

\bibitem[\protect\citeauthoryear{Abdo et al.}{2009}]{2009Sci...325..840A} Abdo A.~A., Ackermann M., Ajello M., Anderson B., Atwood W.~B., Axelsson M., Baldini L., et al., 2009, Sci, 325, 840
\bibitem[\protect\citeauthoryear{Abeysekara et al.}{2017}]{2017Sci...358..911A} Abeysekara A.~U., Albert A., Alfaro R., Alvarez C., {\'A}lvarez J.~D., Arceo R., Arteaga-Vel{\'a}zquez J.~C., et al., 2017, Sci, 358, 911
\bibitem[\protect\citeauthoryear{Acero et al.}{2013}]{2013A&A...551A...7A} Acero F., Gallant Y., Ballet J., Renaud M., Terrier R., 2013, A\&A, 551, A7
\bibitem[\protect\citeauthoryear{Adriani et al.}{2018}]{2018PhRvL.120z1102A} Adriani O., Akaike Y., Asano K., Asaoka Y., Bagliesi M.~G., Berti E., Bigongiari G., et al., 2018, PhRvL, 120, 261102
\bibitem[\protect\citeauthoryear{Aguilar et al.}{2019}]{2019PhRvL.122d1102A} Aguilar M., Ali Cavasonza L., Ambrosi G., Arruda L., Attig N., Azzarello P., Bachlechner A., et al., 2019, PhRvL, 122, 041102
\bibitem[\protect\citeauthoryear{Aguilar et al.}{2014}]{2014PhRvL.113l1102A} Aguilar M., Aisa D., Alvino A., Ambrosi G., Andeen K., Arruda L., Attig N., et al., 2014, PhRvL, 113, 121102
\bibitem[\protect\citeauthoryear{Aharonian, Atoyan, \& Voelk}{1995}]{1995A&A...294L..41A} Aharonian F.~A., Atoyan A.~M., Voelk H.~J., 1995, A\&A, 294, L41
\bibitem[\protect\citeauthoryear{Athron et al.}{2018}]{2018JHEP...02..121A} Athron P., Balazs C., Fowlie A., Zhang Y., 2018, JHEP, 2018, 121
\bibitem[\protect\citeauthoryear{Atoyan, Aharonian, \& V{\"o}lk}{1995}]{1995PhRvD..52.3265A} Atoyan A.~M., Aharonian F.~A., V{\"o}lk H.~J., 1995, PhRvD, 52, 3265
\bibitem[\protect\citeauthoryear{Beck}{2015}]{2015A&ARv..24....4B} Beck R., 2015, A\&ARv,, 24, 4
\bibitem[\protect\citeauthoryear{Belotsky et al.}{2019}]{2019PDU....2600333B} Belotsky K., Kamaletdinov A., Laletin M., Solovyov M., 2019, PDU, 26, 100333
\bibitem[\protect\citeauthoryear{Bergstr{\"o}m, Bringmann, \& Edsj{\"o}}{2008}]{2008PhRvD..78j3520B} Bergstr{\"o}m L., Bringmann T., Edsj{\"o} J., 2008, PhRvD, 78, 103520
\bibitem[\protect\citeauthoryear{Bertone, Hooper, \& Silk}{2005}]{2005PhR...405..279B} Bertone G., Hooper D., Silk J., 2005, PhR, 405, 279
\bibitem[\protect\citeauthoryear{Blasi}{2009}]{2009PhRvL.103e1104B} Blasi P., 2009, PhRvL, 103, 051104
\bibitem[\protect\citeauthoryear{Blasi \& Amato}{2011}]{2011ASSP...21..624B} Blasi P., Amato E., 2011, ASSP, 21, 624
\bibitem[\protect\citeauthoryear{Bucciantini}{2018}]{2018MNRAS.480.5419B} Bucciantini N., 2018, MNRAS, 480, 5419
\bibitem[\protect\citeauthoryear{Caraveo et al.}{2003}]{2003Sci...301.1345C} Caraveo P.~A., Bignami G.~F., De Luca A., Mereghetti S., Pellizzoni A., Mignani R., Tur A., et al., 2003, Sci, 301, 1345
\bibitem[\protect\citeauthoryear{Cerutti \& Philippov}{2017}]{2017A&A...607A.134C} Cerutti B., Philippov A.~A., 2017, A\&A, 607, A134
\bibitem[\protect\citeauthoryear{Chan \& Lee}{2019}]{2019MNRAS.486L..85C} Chan M.~H., Lee C.~M., 2019, MNRAS, 486, L85
\bibitem[\protect\citeauthoryear{Chen, Bandiera, \& Wang}{1996}]{1996ApJ...469..715C} Chen Y., Bandiera R., Wang Z.-R., 1996, ApJ, 469, 715
\bibitem[\protect\citeauthoryear{Coogan, Lehmann, \& Profumo}{2019}]{2019JCAP...10..063C} Coogan A., Lehmann B.~V., Profumo S., 2019, JCAP, 2019, 063
\bibitem[\protect\citeauthoryear{DAMPE Collaboration et al.}{2017}]{2017Natur.552...63D} DAMPE Collaboration, Ambrosi G., An Q., Asfandiyarov R., Azzarello P., Bernardini P., Bertucci B., et al., 2017, Natur, 552, 63
\bibitem[\protect\citeauthoryear{Di Mauro, Manconi, \& Donato}{2020}]{2020PhRvD.101j3035D} Di Mauro M., Manconi S., Donato F., 2020, PhRvD, 101, 103035
\bibitem[\protect\citeauthoryear{Elahi \& Khatibi}{2019}]{2019PhRvD.100a5019E} Elahi F., Khatibi S., 2019, PhRvD, 100, 015019
\bibitem[\protect\citeauthoryear{Fang et al.}{2018}]{2018ApJ...863...30F} Fang K., Bi X.-J., Yin P.-F., Yuan Q., 2018, ApJ, 863, 30
\bibitem[\protect\citeauthoryear{Feng et al.}{2020}]{2020JCAP...04..031F} Feng L., Kang Z., Yuan Q., Yin P.-F., Fan Y.-Z., 2020, JCAP, 2020, 031
\bibitem[\protect\citeauthoryear{Fowlie}{2018}]{2018PhLB..780..181F} Fowlie A., 2018, PhLB, 780, 181
\bibitem[\protect\citeauthoryear{Gabici et al.}{2019}]{2019IJMPD..2830022G} Gabici S., Evoli C., Gaggero D., Lipari P., Mertsch P., Orlando E., Strong A., et al., 2019, IJMPD, 28, 1930022-339
\bibitem[\protect\citeauthoryear{Gaensler \& Slane}{2006}]{2006ARA&A..44...17G} Gaensler B.~M., Slane P.~O., 2006, ARA\&A, 44, 17
\bibitem[\protect\citeauthoryear{Gao \& Ma}{2020}]{2020MNRAS.491..965G} Gao Y., Ma Y.-Z., 2020, MNRAS, 491, 965
\bibitem[\protect\citeauthoryear{H.~E.~S.~S. Collaboration et al.}{2018}]{2018A&A...612A...1H} H.~E.~S.~S. Collaboration, Abdalla H., Abramowski A., Aharonian F., Ait Benkhali F., Ang{\"u}ner E.~O., Arakawa M., et al., 2018, A\&A, 612, A1
\bibitem[\protect\citeauthoryear{Hooper \& Linden}{2018}]{2018PhRvD..98h3009H} Hooper D., Linden T., 2018, PhRvD, 98, 083009
\bibitem[\protect\citeauthoryear{Hooper, Blasi, \& Serpico}{2009}]{2009JCAP...01..025H} Hooper D., Blasi P., Serpico P.~D., 2009, JCAP, 2009, 025
\bibitem[\protect\citeauthoryear{Huang et al.}{2018}]{2018PhRvD..97i1701H} Huang X.-J., Wu Y.-L., Zhang W.-H., Zhou Y.-F., 2018, PhRvD, 97, 091701
\bibitem[\protect\citeauthoryear{Jin et al.}{2018}]{2018PhRvD..98l3008J} Jin H.-B., Yue B., Zhang X., Chen X., 2018, PhRvD, 98, 123008
\bibitem[\protect\citeauthoryear{Li, Lu, \& Li}{2008}]{2008ApJ...682.1166L} Li X.-H., Lu F.-J., Li Z., 2008, ApJ, 682, 1166
\bibitem[\protect\citeauthoryear{Linden \& Profumo}{2013}]{2013ApJ...772...18L} Linden T., Profumo S., 2013, ApJ, 772, 18
\bibitem[\protect\citeauthoryear{Liu \& Liu}{2018}]{2018PhRvD..98c5025L} Liu X., Liu Z., 2018, PhRvD, 98, 035025
\bibitem[\protect\citeauthoryear{Liu, Liu, \& Su}{2019}]{2019JHEP...06..109L} Liu X., Liu Z., Su Y., 2019, JHEP, 2019, 109
\bibitem[\protect\citeauthoryear{L{\'o}pez-Coto et al.}{2018}]{2018APh...102....1L} L{\'o}pez-Coto R., Hahn J., BenZvi S., Dingus B., Hinton J., Nisa M.~U., Parsons R.~D., et al., 2018, APh, 102, 1
\bibitem[\protect\citeauthoryear{Maitra et al.}{2018}]{2018MNRAS.477L..66M} Maitra C., Roy S., Acero F., Gupta Y., 2018, MNRAS, 477, L66
\bibitem[\protect\citeauthoryear{Maitra, Acero, \& Venter}{2017}]{2017A&A...597A..75M} Maitra C., Acero F., Venter C., 2017, A\&A, 597, A75
\bibitem[\protect\citeauthoryear{Malyshev, Cholis, \& Gelfand}{2009}]{2009PhRvD..80f3005M} Malyshev D., Cholis I., Gelfand J., 2009, PhRvD, 80, 063005
\bibitem[\protect\citeauthoryear{Manchester et al.}{2005}]{2005AJ....129.1993M} Manchester R.~N., Hobbs G.~B., Teoh A., Hobbs M., 2005, AJ, 129, 1993
\bibitem[\protect\citeauthoryear{Ng et al.}{2017}]{2017ApJ...842..100N} Ng C.-Y., Bandiera R., Hunstead R.~W., Johnston S., 2017, ApJ, 842, 100
\bibitem[\protect\citeauthoryear{Nomura, Okada, \& Wu}{2018}]{2018JCAP...05..053N} Nomura T., Okada H., Wu P., 2018, JCAP, 2018, 053
\bibitem[\protect\citeauthoryear{Olmi \& Bucciantini}{2019}]{2019MNRAS.490.3608O} Olmi B., Bucciantini N., 2019, MNRAS, 490, 3608
\bibitem[\protect\citeauthoryear{Posselt et al.}{2017}]{2017ApJ...835...66P} Posselt B., Pavlov G.~G., Slane P.~O., Romani R., Bucciantini N., Bykov A.~M., Kargaltsev O., et al., 2017, ApJ, 835, 66
\bibitem[\protect\citeauthoryear{Profumo et al.}{2018}]{2018PhRvD..97l3008P} Profumo S., Reynoso-Cordova J., Kaaz N., Silverman M., 2018, PhRvD, 97, 123008
\bibitem[\protect\citeauthoryear{Schlickeiser}{2002}]{2002cra..book.....S} Schlickeiser R., 2002, cra..book
\bibitem[\protect\citeauthoryear{Shaviv, Nakar, \& Piran}{2009}]{2009PhRvL.103k1302S} Shaviv N.~J., Nakar E., Piran T., 2009, PhRvL, 103, 111302
\bibitem[\protect\citeauthoryear{Shen}{1970}]{1970ApJ...162L.181S} Shen C.~S., 1970, ApJL, 162, L181
\bibitem[\protect\citeauthoryear{Torres et al.}{2014}]{2014JHEAp...1...31T} Torres D.~F., Cillis A., Mart{\'\i}n J., de O{\~n}a Wilhelmi E., 2014, JHEAp, 1, 31
\bibitem[\protect\citeauthoryear{Volwerk \& Kuijpers}{1994}]{1994SSRv...68..363V} Volwerk M., Kuijpers J., 1994, SSRv, 68, 363
\bibitem[\protect\citeauthoryear{Yuan et al.}{2017}]{2017PhRvD..95h3007Y} Yuan Q., Lin S.-J., Fang K., Bi X.-J., 2017, PhRvD, 95, 083007


\end{thebibliography}




\appendix

\bsp	
\label{lastpage}
\end{document}